\documentclass[lettersize,journal]{IEEEtran}
\IEEEoverridecommandlockouts
\usepackage{cite}
\usepackage{amsmath,amssymb,amsfonts}

\usepackage{algorithm}
\usepackage{algpseudocode}
\usepackage{algorithmicx}
\usepackage{amsmath}
\usepackage{xcolor}

\newcommand{\Phase}[1]{\Statex \textbf{\textcolor{black}{#1:}}\Statex}

\usepackage{graphicx}
\usepackage{multirow}
\usepackage{textcomp}
\usepackage{xcolor}
\usepackage{algorithm}
\usepackage{algpseudocode}

\def\BibTeX{{\rm B\kern-.05em{\sc i\kern-.025em b}\kern-.08em
    T\kern-.1667em\lower.7ex\hbox{E}\kern-.125emX}}
\usepackage{multirow}
\usepackage{algorithm}
\usepackage{array}
\usepackage{geometry}
\usepackage[caption=false,font=normalsize,labelfont=sf,textfont=sf]{subfig}
\usepackage{textcomp}
\usepackage{stfloats}
\usepackage{url}
\usepackage{xcolor}
\usepackage{verbatim}
\usepackage{caption} 
\usepackage{graphicx}
\usepackage{subfig}
\usepackage{enumerate}
\usepackage{enumitem}
\usepackage{tabularx,booktabs,textcomp}
\usepackage{amsthm}

\usepackage{amsmath,amssymb,amsthm}

\makeatletter 
\newcommand{\linebreakand}{%
  \end{@IEEEauthorhalign}
  \hfill\mbox{}\par
  \mbox{}\hfill\begin{@IEEEauthorhalign}
}

\begin{document}

\title{Integrated Optimization and Game Theory Framework for Fair Cost Allocation in Community Microgrids}

\author{{K. Victor Sam Moses Babu,~\IEEEmembership{Member,~IEEE},  Pratyush Chakraborty,~\IEEEmembership{Senior Member,~IEEE}, Mayukha Pal,~\IEEEmembership{Senior Member,~IEEE}}

\thanks{(Corresponding author: Mayukha Pal)}

\thanks{Mr. K. Victor Sam Moses Babu is a Data Science Research Intern at ABB Ability Innovation Center, Hyderabad 500084, India and also a Research Scholar at the Department of Electrical and Electronics Engineering, BITS Pilani Hyderabad Campus, Hyderabad 500078, IN.}
\thanks{Dr. Pratyush Chakraborty is an Asst. Professor with the Department of Electrical and Electronics Engineering, BITS Pilani Hyderabad Campus, Hyderabad 500078, IN.}
\thanks{Dr. Mayukha Pal is with ABB Ability Innovation Center, Hyderabad-500084, IN, working as Global R\&D Leader – Cloud \& Analytics (e-mail: mayukha.pal@in.abb.com).}
}
\maketitle

\begin{abstract}
Fair cost allocation in community microgrids remains a significant challenge due to the complex interactions between multiple participants with varying load profiles, distributed energy resources, and storage systems. Traditional cost allocation methods often fail to adequately address the dynamic nature of participant contributions and benefits, leading to inequitable distribution of costs and reduced participant satisfaction. This paper presents a novel framework integrating multi-objective optimization with cooperative game theory for fair and efficient microgrid operation and cost allocation. The proposed approach combines mixed-integer linear programming for optimal resource dispatch with Shapley value analysis for equitable benefit distribution, ensuring both system efficiency and participant satisfaction. The framework was validated using real-world data across six distinct operational scenarios, demonstrating significant improvements in both technical and economic performance. Results show peak demand reductions ranging from 7.8\% to 62.6\%, solar utilization rates reaching 114.8\% through effective storage integration, and cooperative gains of up to \$1,801.01 per day. The Shapley value-based allocation achieved balanced benefit-cost distributions, with net positions ranging from -16.0\% to +14.2\% across different load categories, ensuring sustainable participant cooperation.
\end{abstract}


\section{Introduction}

The transition towards sustainable energy systems has led to increased adoption of community microgrids, which offer enhanced reliability, improved energy efficiency, and greater integration of renewable resources \cite{wang2022transactive, salehi2022comprehensive, DWIVEDI2024110537}. These systems represent a paradigm shift from traditional centralized power distribution to collaborative energy management, where multiple stakeholders actively participate in both generation and consumption of energy \cite{billah2023decentralized}. The proliferation of distributed energy resources (DERs) and energy storage systems in community microgrids creates complex operational dynamics that traditional management approaches struggle to address effectively \cite{akarne2024experimental, yuan2023data}. These systems must balance multiple, often competing objectives: maximizing renewable energy utilization, minimizing operational costs, reducing peak demand, and ensuring reliable power supply \cite{wang2022transactive, billah2023decentralized}. The intermittent nature of renewable generation and varying load patterns further complicate this balancing act, necessitating sophisticated optimization approaches \cite{akarne2024experimental}.

Microgrid optimization represents a critical challenge in modern power systems, where the interaction between multiple energy sources, storage systems, and loads must be carefully coordinated \cite{salehi2022comprehensive, yuan2023data}. The optimization problem encompasses various technical constraints, including power balance, battery operational limits, and grid interface requirements \cite{wang2022transactive}, while considering economic factors such as time-varying electricity prices and demand charges \cite{billah2023decentralized, akarne2024experimental}. The complexity of microgrid operations is further amplified by the need to consider multiple time scales, from real-time power balance to daily and seasonal variations in generation and demand \cite{yuan2023data, wang2022_ref31}. Energy storage systems play a crucial role in this context, enabling temporal shifting of energy and peak demand management \cite{akarne2024experimental, faisal2018review}. However, the optimal utilization of storage resources requires careful consideration of degradation costs and operational constraints \cite{sinha2020power}, adding another layer of complexity to the optimization problem \cite{bandeiras2020review, 10000399}.

Existing literature has proposed various approaches to microgrid optimization and cost allocation. Traditional methods often employ rule-based or heuristic approaches \cite{hannan2020review}, which may fail to achieve optimal system performance \cite{K2024109502}. More advanced techniques include model predictive control \cite{hu2021model} and multi-objective optimization \cite{salehi2022comprehensive}, but these typically focus on technical performance metrics without addressing the fairness of cost allocation. Game theory approaches have been explored for cost allocation \cite{yuan2023data, churkin2021review}, but most existing work treats optimization and allocation as separate problems, potentially leading to suboptimal solutions.
Recent research has investigated the application of cooperative game theory to microgrid cost allocation \cite{abdulmohsen2022active, 10636225}, with some studies utilizing Shapley value analysis to determine fair cost distributions \cite{wang2022transactive}. However, these approaches often consider simplified system models or fail to integrate the allocation mechanism with the operational optimization \cite{alam2019networked}, limiting their practical applicability. Additionally, most existing work lacks comprehensive validation across diverse operational scenarios \cite{jadhav2019novel, li2021impact}.
The integration of demand response and renewable energy sources introduces additional complexities in microgrid management \cite{yuan2023data, prajapati2022demand}. These systems must handle uncertainties in both generation and consumption patterns \cite{jiao2020multi}, while maintaining system stability and power quality \cite{ramadan2023optimal}. Advanced control strategies, such as Multi-Agent Systems (MAS) \cite{salehi2022comprehensive} and hierarchical control architectures \cite{billah2023decentralized}, have been proposed to address these challenges, but their integration with fair cost allocation mechanisms remains an open research question.

\begin{figure}[t]
\centering
\includegraphics[width=3.5in]{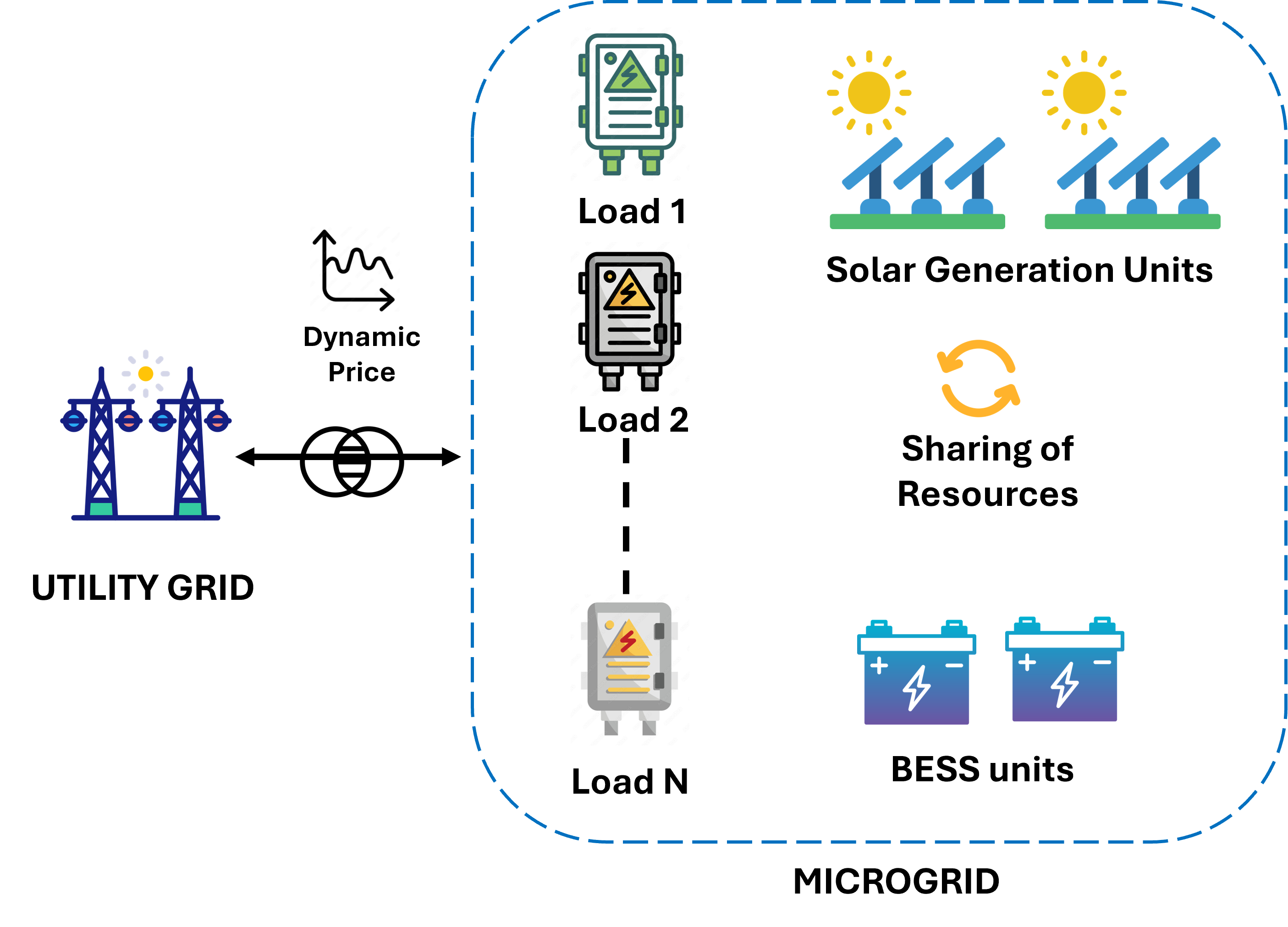}
  \caption{Schematic of a grid-connected community microgrid.}
  \label{fig:microgrid_architecture}
\end{figure}

The proposed framework addresses these limitations by integrating mixed-integer linear programming optimization with Shapley value-based cost allocation in a unified solution \cite{wang2022transactive, kharrich2021multi}. As shown in Fig. \ref{fig:microgrid_architecture}, the framework is implemented on a community microgrid comprising multiple loads, distributed solar generation units, and shared battery energy storage systems. The microgrid operates under dynamic pricing from the utility grid, with participants sharing the distributed energy resources and storage facilities. This integration ensures simultaneous optimization of both system efficiency and fairness objectives through a two-stage solution approach \cite{salehi2022comprehensive}. The first stage employs MILP to optimize the microgrid operation considering battery degradation costs, peak demand charges, and grid connection costs, while the second stage utilizes Shapley value analysis for equitable benefit distribution \cite{akarne2024experimental, liu2020multi}. The framework maintains computational tractability through strategic problem decomposition and parallel processing of coalition evaluations, enabling practical implementation for real-world applications. The effectiveness of this approach is validated across six distinct operational scenarios: peak demand day, low demand day, high price day, high solar generation day, typical weekday, and typical weekend, demonstrating robust performance under diverse conditions.

The key contributions of this work include: 
\begin{enumerate}
    \item Development of a unified framework integrating operational optimization with fair cost allocation.
    \item Implementation of a mixed-integer linear programming model capturing all relevant system constraints and interactions. 
    \item Application of Shapley value analysis for transparent and equitable benefit distribution.
    \item Comprehensive validation across six distinct operational scenarios with detailed analysis of technical and economic performance metrics. 
\end{enumerate}

The remainder of this paper is organized as follows: Section \ref{ch:methods} presents the detailed methodology, including the optimization model and Shapley value analysis. Section \ref{ch:simulation} describes the simulation setup and discusses the results, including technical performance, economic metrics, and fairness analysis. Finally, Section \ref{ch:conclusion} provides conclusions and future research directions.

\section{Materials and Methods}
\label{ch:methods}

\section{Methodology}

This section presents a comprehensive framework for optimizing microgrid operations while ensuring fair cost allocation among participants. The proposed methodology addresses three critical challenges in microgrid management: optimal resource utilization, equitable cost distribution, and sustainable cooperative operation. To tackle these challenges, we integrate three main components: (1) multi-objective mixed-integer linear programming (MILP) for microgrid operation optimization, (2) cooperative game theory using Shapley value analysis for benefit allocation, and (3) enhanced fairness analysis for participant equity assessment.

\subsection{Microgrid System Architecture}

The microgrid system under consideration represents a complex energy ecosystem comprising multiple stakeholders and resources. The system includes $N$ loads representing different consumers (e.g., residential, commercial), $S$ distributed solar generation units providing renewable energy, and $B$ battery energy storage systems (BESS) enabling temporal energy shifting. The system operates over $T$ discrete time periods (24 hours) with the capability to import power from the main grid when needed.

Each participant $i \in \{1,\ldots,N\}$ has a unique time-varying load profile $L_i(t)$ for $t \in \{1,\ldots,T\}$, reflecting their individual energy consumption patterns. The solar generation units provide power $S_j(t)$ for $j \in \{1,\ldots,S\}$, which varies based on weather conditions and time of day. Each BESS unit $k \in \{1,\ldots,B\}$ has specific operational characteristics including capacity, charging/discharging rates, and efficiency parameters.

\subsection{Mixed-Integer Linear Programming Optimization}

The optimization framework aims to minimize operational costs while satisfying technical constraints and ensuring reliable power supply. We formulate this as a MILP problem to capture both continuous power flows and discrete operational decisions.

\subsubsection{Decision Variables}

The optimization problem considers several categories of decision variables, each serving a specific purpose in the system operation:

\begin{itemize}
    \item Grid Power Import ($G_i(t)$): Represents the power drawn from the main grid for each load $i$ at time $t$. This variable is crucial for cost minimization as it directly impacts both energy costs and peak demand charges.
    
    \item Solar Allocation ($\alpha_{ij}(t)$): Determines the fraction of solar power from unit $j$ allocated to load $i$ at time $t$. This variable enables fair distribution of renewable resources among participants.
    
    \item BESS Operation:
    \begin{itemize}
        \item Charging Power ($C_{ik}(t)$): Power flowing from the grid or solar units to BESS $k$ through load connection point $i$ at time $t$
        \item Discharging Power ($D_{ik}(t)$): Power supplied from BESS $k$ to load $i$ at time $t$
        \item State of Charge ($SOC_k(t)$): Energy level of BESS $k$ at time $t$
    \end{itemize}
    
    \item System Peak ($P_{peak}$): Maximum power drawn from the grid across all time periods, which directly affects demand charges
\end{itemize}

\subsubsection{Objective Function}

The optimization minimizes the total operational cost, which comprises two main components:

\begin{equation}
\min \sum\limits\limits_{t=1}^T \sum\limits\limits_{i=1}^N p(t)G_i(t) + C_p \cdot P_{peak}
\end{equation}

The first term represents the energy costs, where $p(t)$ is the time-varying electricity price. This captures the temporal variations in grid electricity costs and encourages consumption during lower-price periods. The second term accounts for peak demand charges, where $C_p$ is the peak demand charge rate (set to \$8,700/MW). This substantial demand charge creates a strong incentive for peak shaving and load shifting.

\subsubsection{Power Balance Constraints}

The fundamental requirement for reliable microgrid operation is maintaining power balance at all times. For each load and time period, the sum of power from all sources must meet the demand:

\begin{equation}
G_i(t) + \sum\limits\limits_{j=1}^S \alpha_{ij}(t)S_j(t) + \sum\limits\limits_{k=1}^B [D_{ik}(t) - C_{ik}(t)] = L_i(t)(1 \pm \epsilon)
\end{equation}

Here, we introduce a small tolerance $\epsilon$ (0.1
- Grid import ($G_i(t)$)
- Allocated solar generation ($\sum\limits\limits_{j=1}^S \alpha_{ij}(t)S_j(t)$)
- Net battery power ($\sum\limits\limits_{k=1}^B [D_{ik}(t) - C_{ik}(t)]$)

This constraint ensures that each participant's load is met while allowing flexibility in the source of power.

\subsubsection{Solar Generation Constraints}

The solar power allocation must respect physical and operational limits:

\begin{equation}
\sum\limits\limits_{i=1}^N \alpha_{ij}(t) \leq 1, \quad \forall j,t
\end{equation}

This constraint ensures that the total allocation of each solar unit's output doesn't exceed 100\%. Additionally, we enforce non-negative solar allocation:

\begin{equation}
\alpha_{ij}(t)S_j(t) \geq 0, \quad \forall i,j,t
\end{equation}

These constraints prevent overselling of solar generation and ensure realistic power flows.

\subsubsection{BESS Operational Constraints}

Battery energy storage systems play a crucial role in temporal energy arbitrage and peak shaving. Their operation is governed by several complex constraints to ensure practical and efficient operation.

1. State of Charge (SOC) Evolution:
The SOC of each battery evolves based on charging and discharging activities:

\begin{equation}
SOC_k(t) = SOC_k(t-1) + \eta_c \sum\limits\limits_{i=1}^N C_{ik}(t) - \dfrac{1}{\eta_d} \sum\limits\limits_{i=1}^N D_{ik}(t)
\end{equation}

where $\eta_c$ and $\eta_d$ are charging and discharging efficiencies respectively. This equation captures energy losses during battery operation and ensures energy conservation. The SOC at each time step depends on the previous state and net energy flow, accounting for efficiency losses in both charging and discharging processes.

2. SOC Limits:
To preserve battery life and ensure reliable operation, the SOC must remain within specified bounds:

\begin{equation}
SOC_k^{min}E_k \leq SOC_k(t) \leq SOC_k^{max}E_k
\end{equation}

where $E_k$ is the battery capacity, $SOC_k^{min}$ = 0.15 and $SOC_k^{max}$ = 0.95. These limits prevent deep discharge and overcharging, which can significantly impact battery longevity. The minimum SOC of 15\% provides a safety margin for unexpected demand spikes, while the maximum of 95

3. Power Limits:
The charging and discharging rates are constrained by the battery's power capability:

\begin{equation}
\sum\limits_{i=1}^N C_{ik}(t) \leq P_k^{max}
\end{equation}

\begin{equation}
\sum\limits_{i=1}^N D_{ik}(t) \leq P_k^{max}
\end{equation}

These constraints reflect the physical limitations of power conversion systems and protect the battery from excessive power flows that could cause degradation or safety issues.

4. Prevention of Simultaneous Charging and Discharging:
To prevent inefficient operation and potential damage:

\begin{equation}
\sum\limits_{i=1}^N C_{ik}(t) + \sum\limits_{i=1}^N D_{ik}(t) \leq P_k^{max}
\end{equation}

This constraint eliminates the possibility of simultaneous charging and discharging, which would result in unnecessary energy losses and potential stability issues.

5. Terminal SOC Requirement:
To ensure system readiness for the next operational cycle:

\begin{equation}
SOC_k(T) \geq 0.4E_k
\end{equation}

This constraint maintains a reasonable energy reserve (40\% of capacity) at the end of the optimization period, ensuring availability for the next day's operation and providing resilience against unexpected events.

6. Minimum Up/Down Time:
To prevent excessive cycling and protect battery life, we enforce minimum duration for charging and discharging operations. For any time $t$ where operation begins:

\begin{equation}
\sum\limits_{i=1}^N \sum\limits_{t'=t}^{t+\tau_{min}} C_{ik}(t') \geq 0
\end{equation}

\begin{equation}
\sum\limits_{i=1}^N \sum\limits_{t'=t}^{t+\tau_{min}} D_{ik}(t') \geq 0
\end{equation}

where $\tau_{min}$ is set to 2 hours. This prevents rapid switching between charging and discharging states, which can accelerate battery degradation and reduce system efficiency.

\subsubsection{Grid Interface Constraints}

The interaction with the main grid must be carefully managed to minimize costs and ensure stable operation.

1. Peak Demand Management:
The system's peak power draw is constrained and tracked:

\begin{equation}
\sum\limits_{i=1}^N G_i(t) \leq P_{peak}, \quad \forall t
\end{equation}

This constraint defines the peak demand value used in cost calculations while ensuring that total grid import never exceeds the declared peak power.

2. Ramp Rate Limits:
To maintain grid stability and prevent sudden large power variations:

\begin{equation}
|G_i(t) - G_i(t-1)| \leq R_{max} \cdot L_i(t)
\end{equation}

where $R_{max}$ = 0.2, limiting ramp rates to 20\% of the load per hour. This constraint promotes smooth operation and helps prevent grid instability while also protecting equipment from rapid power fluctuations.

\subsection{Shapley Value-Based Cost Allocation}

After optimizing the microgrid operation, a crucial challenge is fairly distributing the costs and benefits among participants. We employ Shapley value analysis, a cooperative game theory concept, to ensure fair and stable cost allocation that incentivizes continued participation.

The Shapley value $\phi_i$ for each participant $i$ is calculated as:

\begin{equation}
\phi_i = \sum\limits_{S \subseteq N \setminus \{i\}} \dfrac{|S|!(|N|-|S|-1)!}{|N|!}[v(S \cup \{i\}) - v(S)]
\end{equation}

This formulation considers all possible coalitions $S$ and calculates each participant's marginal contribution. The factorial terms ensure that the order of joining the coalition doesn't affect the allocation. We calculate separate Shapley values for four key components to provide a comprehensive and transparent allocation mechanism.

\subsubsection{Solar Benefits Distribution}

The solar benefit allocation considers the value of renewable energy utilized by each participant:

\begin{equation}
v_s(S) = \sum\limits_{t=1}^T \sum\limits_{j=1}^S \sum\limits_{i \in S} p(t)\alpha_{ij}(t)S_j(t)
\end{equation}

This characteristic function values solar generation at the prevailing grid price $p(t)$, reflecting the actual cost savings from reduced grid imports. The allocation considers both the quantity of solar power received ($\alpha_{ij}(t)S_j(t)$) and its time-of-use value.

\subsubsection{BESS Costs Allocation}

Battery costs are allocated based on utilization patterns:

\begin{equation}
v_b(S) = C_b \sum\limits_{t=1}^T \sum\limits_{k=1}^B \sum\limits_{i \in S} (C_{ik}(t) + D_{ik}(t))
\end{equation}

where $C_b$ = \$10/MWh represents the battery degradation cost. This allocation ensures that participants who make greater use of battery services contribute proportionally to the wear and tear costs. Both charging and discharging actions are considered as they both impact battery lifetime.

\subsubsection{Peak Savings Distribution}

Peak demand reduction benefits are particularly significant due to high demand charges:

\begin{equation}
v_p(S) = C_p(\max_{t} \sum\limits_{i \in S} L_i(t) - \max_{t} \sum\limits_{i \in S} G_i(t))
\end{equation}

This formulation quantifies the savings from peak reduction by comparing the original load peak to the optimized grid import peak, valued at the demand charge rate $C_p$. Participants who contribute more to peak reduction receive larger shares of these savings.

\subsubsection{Grid Infrastructure Cost Allocation}

Fixed grid costs are distributed based on relative grid utilization:

\begin{equation}
v_g(S) = C_f \cdot \dfrac{\sum\limits_{t=1}^T \sum\limits_{i \in S} G_i(t)}{\sum\limits_{t=1}^T \sum\limits_{i=1}^N G_i(t)}
\end{equation}

where $C_f$ = \$1,000 represents the daily fixed infrastructure cost. This approach ensures that participants who rely more heavily on grid imports bear a proportionally larger share of the fixed costs.

\subsection{Enhanced Fairness Analysis}

To ensure the long-term sustainability of the cooperative arrangement, we implement comprehensive fairness metrics that evaluate the equity and stability of the allocation.

\subsubsection{Net Position Analysis}

The net position ($NP_i$) of each participant balances their benefits against costs:

\begin{equation}
NP_i = \dfrac{\phi_i^s + \phi_i^p}{\sum\limits_{j=1}^N (\phi_j^s + \phi_j^p)} - \dfrac{\phi_i^b + \phi_i^g}{\sum\limits_{j=1}^N (\phi_j^b + \phi_j^g)}
\end{equation}

This metric provides a normalized measure of each participant's overall financial position, considering both received benefits (solar and peak savings) and incurred costs (battery and grid charges). A positive net position indicates a participant receiving more benefits than costs relative to their size and usage.

\subsubsection{Cooperative Gains Analysis}

The total benefit of cooperation is quantified as:

\begin{equation}
CG = \sum\limits_{i=1}^N C_i^{ind} - C_{total}^{coop}
\end{equation}

where $C_i^{ind}$ represents the cost each participant would incur operating independently, and $C_{total}^{coop}$ is the total cost under cooperative operation. This metric demonstrates the value of participation and helps justify the cooperative arrangement.

The average gain per participant is calculated as:

\begin{equation}
AG = \dfrac{CG}{N}
\end{equation}

This metric helps in communicating the benefits of cooperation to stakeholders and provides a benchmark for evaluating the success of the arrangement.

\subsubsection{Load Proportionality Analysis}

To assess fairness relative to system usage, we calculate load proportions:

\begin{equation}
LP_i = \dfrac{\sum\limits_{t=1}^T L_i(t)}{\sum\limits_{t=1}^T \sum\limits_{j=1}^N L_j(t)}
\end{equation}

The deviation from proportional allocation is then analyzed:

\begin{equation}
\delta_i = |NP_i - LP_i|
\end{equation}

This analysis helps identify any systematic biases in the allocation mechanism and ensures that benefits and costs are reasonably aligned with system usage.

\subsubsection{Synergy Metrics}

We evaluate system performance through several key metrics:

1. Solar Utilization Rate:
\begin{equation}
SUR = \dfrac{\sum\limits_{t=1}^T \sum\limits_{i=1}^N \sum\limits_{j=1}^S \alpha_{ij}(t)S_j(t)}{\sum\limits_{t=1}^T \sum\limits_{j=1}^S S_j(t)}
\end{equation}

This metric quantifies the efficiency of solar resource utilization, with higher values indicating better use of available renewable energy.

2. BESS Cycling Rate:
\begin{equation}
BCR_k = \dfrac{\sum\limits_{t=1}^T \sum\limits_{i=1}^N D_{ik}(t)}{E_k}
\end{equation}

This measures the utilization intensity of each battery, helping in maintenance planning and lifetime estimation.

3. Peak Reduction Efficiency:
\begin{equation}
PRE = \dfrac{\max_{t} \sum\limits_{i=1}^N L_i(t) - P_{peak}}{\max_{t} \sum\limits_{i=1}^N L_i(t)}
\end{equation}

This metric quantifies the effectiveness of the system in reducing peak demand, directly impacting cost savings.

This comprehensive methodology, presented in Algorithm \ref{alg:framework}, provides a robust framework for optimal microgrid operation and equitable benefit distribution. The algorithm consists of four key phases: (1) microgrid operation optimization using MILP, (2) cooperative game theory analysis through Shapley value calculations, (3) comprehensive fairness analysis with multiple performance metrics, and (4) results validation and visualization. The integration of detailed operational constraints with sophisticated cost allocation mechanisms ensures practical feasibility while maintaining participant satisfaction and system sustainability.

\begin{algorithm}
\caption{Microgrid Optimization and Fair Cost Allocation Framework}
\label{alg:framework}
\begin{algorithmic}[1]
\Require
    \State Load profiles $L_i(t)$ for each participant
    \State Solar generation profiles $S_j(t)$
    \State Battery specifications (capacity, efficiency, power limits)
    \State Time-of-use electricity prices $p(t)$
    \State Peak demand charge rate $C_p$

\Ensure
    \State Optimal operation schedule
    \State Fair cost and benefit allocation
    \State Performance metrics

\Phase{1: Microgrid Operation Optimization}
    \State Initialize MILP optimization model
    \State Define decision variables (grid import, solar allocation, battery operation)
    \State Set objective function (minimize energy cost + peak charges)
    \State Add power balance constraints
    \State Add battery operational constraints
    \State Add solar allocation constraints
    \State Solve optimization problem using CBC solver
    \State Store optimal operation schedule
    \State Calculate total system costs

\Phase{2: Cooperative Game Theory Analysis}
    \State Initialize Shapley value calculations
    \For{each participant $i$}
        \State Calculate individual operation cost
        \For{each possible coalition $S$}
            \State Calculate coalition benefits:
            \State - Solar generation benefits
            \State - Peak reduction savings
            \State - Battery utilization costs
            \State - Grid infrastructure costs
        \EndFor
        \State Compute Shapley values for each component
    \EndFor

\Phase{3: Fairness Analysis and Performance Evaluation}
    \For{each participant $i$}
        \State Calculate net position
        \State Compute load proportion
        \State Determine benefit-to-contribution ratio
    \EndFor
    \State Calculate system-wide metrics:
    \State - Cooperative gains
    \State - Solar utilization rate
    \State - Battery cycling efficiency
    \State - Peak reduction effectiveness
    
\Phase{4: Results Compilation and Validation}
    \State Generate cost allocation report
    \State Verify fairness criteria
    \State Compute performance metrics
    \State Validate stability of allocation
    \State Prepare visualization of results

\Return Optimal schedule, cost allocation, and performance metrics
\end{algorithmic}
\end{algorithm}

\section{Simulation and Results}
\label{ch:simulation}

\subsection{Simulation Setup and Data Processing}

The microgrid configuration consists of diverse load profiles representing different types of consumers. The loads exhibit varying consumption patterns with peak demands ranging from 0.15 MW to 0.25 MW per consumer. Two solar PV systems, each rated at 1.5 MW peak capacity, provide renewable generation. The battery energy storage systems each have a capacity of 1.0 MWh with maximum charging/discharging rates of 0.5 MW. The battery systems operate with round-trip efficiency of 92\%, consisting of 96\% charging efficiency and 96\% discharging efficiency. Time-series data \cite{Dataset} was collected at hourly intervals (24 points per day) for six distinct scenarios representing different operational conditions: peak demand day, low demand day, high price day, high solar generation day, typical weekday, and typical weekend. The electricity price data follows the local utility's time-of-use tariff structure, varying between \$68.90/MWh and \$339.91/MWh. A substantial peak demand charge of \$8,700/MW-month is applied to incentivize peak reduction. 

\subsection{Implementation Details}

Thus, the proposed microgrid optimization and fair cost allocation framework is implemented and tested using real-world data from a community microgrid system comprising ten loads, two distributed solar generation units, and two battery energy storage systems. The simulation environment was developed utilizing PuLP 2.4 for optimization modeling and CBC (COIN-OR Branch and Cut) solver for solving the mixed-integer linear programming problem. A tolerance of 0.1\% was introduced in the power balance constraints to improve solver convergence. The CBC solver was configured with a time limit of 300 seconds and an optimality gap tolerance of 0.5\% to ensure practical solution times while maintaining solution quality.

Battery operation constraints were implemented with particular attention to lifecycle considerations. The state of charge limits were set to 15\% minimum and 95\% maximum, with a required terminal state of charge of 40\% to ensure system availability for the next operational cycle. The minimum up/down time of 2 hours for battery operation was enforced to prevent excessive cycling, while ramp rate limits of 20\% per hour were applied to grid power variations to ensure stable operation. The Shapley value calculations for cost allocation were implemented using a parallel processing approach to handle the computational complexity of evaluating all possible coalitions. For a system with ten participants, this involved analyzing 1,024 different coalition combinations for each cost component. The characteristic functions for solar benefits, battery costs, peak savings, and grid costs were computed using vectorized operations to improve computational efficiency.

For a comprehensive evaluation of the framework's performance, several key metrics were tracked across all scenarios. Economic metrics include baseline energy costs, optimized energy costs, peak demand charges, and total system costs. Technical performance metrics encompass peak demand reduction, solar utilization rate, battery cycling, and overall system efficiency. The fairness of cost allocation was evaluated through load proportionality analysis, benefit distribution metrics, and cooperative gain calculations. Table \ref{tab:system_params} presents the key system parameters used in the simulation.

\begin{table}[h!]
\centering
\caption{System Parameters and Operational Limits}
\label{tab:system_params}
\begin{tabular}{ll}
\hline
Parameter & Value \\
\hline
Number of Loads & 10 \\
Number of Solar Units & 2 (1.5 MW each) \\
Number of BESS Units & 2 (1.0 MWh each) \\
BESS Power Rating & 0.5 MW \\
BESS Round-trip Efficiency & 92\% \\
Minimum SOC & 15\% \\
Maximum SOC & 95\% \\
Terminal SOC Requirement & 40\% \\
Grid Ramp Rate Limit & 20\% per hour \\
Peak Demand Charge & \$8,700/MW \\
Battery Degradation Cost & \$10/MWh \\
Grid Connection Cost & \$1,000/day \\
Optimization Time Limit & 300 seconds \\
Power Balance Tolerance & 0.1\% \\
\hline
\end{tabular}
\end{table}

\subsection{Scenario Analysis and Results}

The framework was evaluated across six distinct scenarios to demonstrate its effectiveness under different operating conditions. Each scenario presents unique challenges and opportunities for the microgrid operation and cost allocation system.

\subsubsection{Peak Demand Day Analysis (December 30, 2014)}

\begin{figure}[h!]
\centering
\includegraphics[width=3.5in]{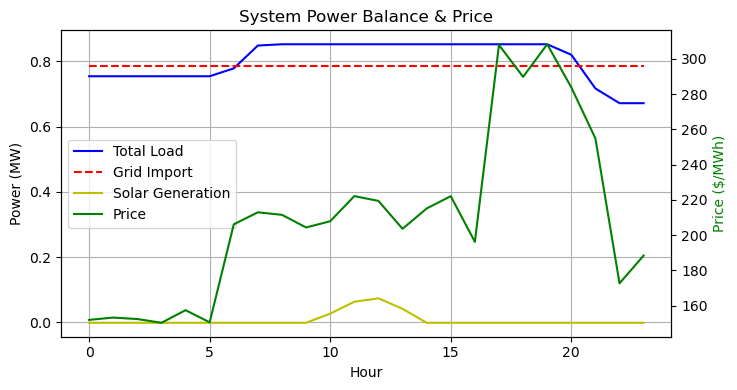}
  \caption{System power and grid price - Peak Demand Day.}
  \label{fig:S1}
\end{figure}

The peak demand day scenario represents the most challenging operational conditions for the microgrid. The original peak load reached 0.853 MW, with significant variation in demand patterns across the ten loads. The optimization framework successfully reduced the peak demand to 0.786 MW, as shown in Fig. \ref{fig:S1}, achieving a 7.8\% reduction through coordinated battery dispatch and solar utilization.

\begin{figure}[h!]
\centering
\includegraphics[width=3.5in]{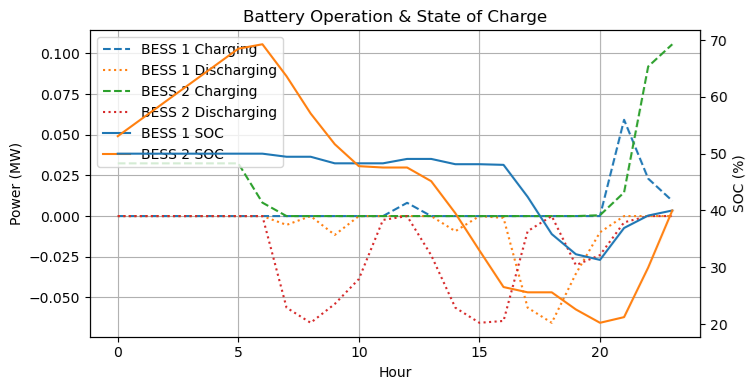}
  \caption{Battery operation and SoC - Peak Demand Day.}
  \label{fig:S1_-_b}
\end{figure}
\begin{figure}[h!]
\centering
\includegraphics[width=3.5in]{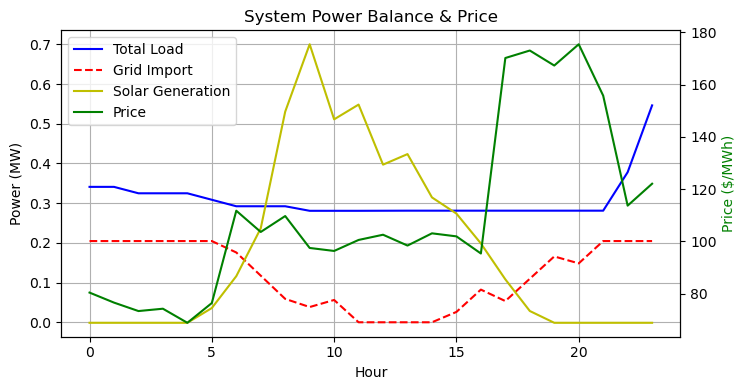}
  \caption{System power and grid price - Low Demand Day.}
  \label{fig:S2}
\end{figure}
\begin{figure}[h!]
\centering
\includegraphics[width=3.5in]{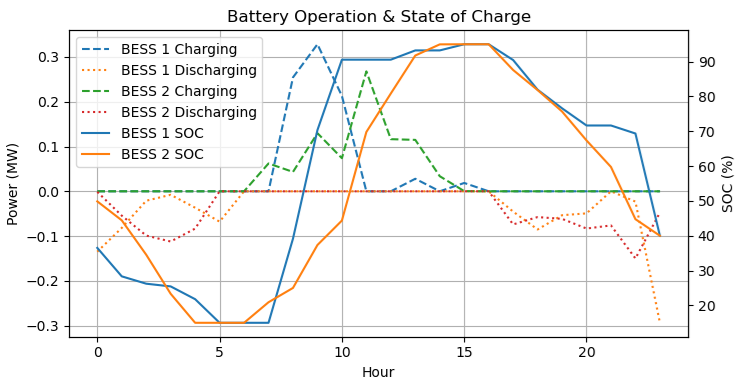}
  \caption{Battery operation and SoC - Low Demand Day.}
  \label{fig:S2_-_b}
\end{figure}

The economic analysis reveals substantial cost savings through optimization. The baseline energy cost of \$4,084.46 was reduced to \$3,964.31 through optimal resource allocation. Including the peak demand charge of \$6,840.90, the total optimized cost was \$10,805.21. Battery utilization showed conservative cycling patterns, with BESS 1 completing 0.19 cycles and BESS 2 completing 0.49 cycles, discharging 0.19 MWh and 0.49 MWh respectively. Both batteries maintained healthy SOC ranges, , as shown in Fig. \ref{fig:S1_-_b}, with BESS 1 operating between 31.3\% and 50.0\%, and BESS 2 between 20.2\% and 69.3\%.

The solar generation was fully utilized with a 114.8\% utilization rate, indicating effective temporal shifting of solar energy through battery storage. Individual load analysis showed varied benefits, with Load 8 receiving the highest solar allocation (0.05 MWh) and Load 9 the lowest (0.00 MWh). Battery support varied significantly, ranging from -0.04 MWh (net charging) for Load 8 to 0.12 MWh (net discharging) for Load 4.
Shapley value analysis revealed interesting patterns in benefit distribution. Solar benefits were highest for Load 8 (22.2\%) and Load 2 (21.7\%), while Load 9 received no solar benefits. BESS costs were predominantly borne by Load 5 (16.9\%) and Load 9 (11.8\%). Peak savings were relatively evenly distributed, ranging from 8.9\% to 11.9\%. The cooperative game analysis showed significant benefits of cooperation, with total cooperative gains of \$1,615.72, averaging \$161.57 per participant.

\subsubsection{Low Demand Day Analysis (August 15, 2015)}

The low demand day presented opportunities for maximizing renewable utilization and minimizing grid dependence. The optimization achieved remarkable peak reduction of 62.6\%, reducing the original peak load of 0.546 MW to 0.205 MW, as shown in Fig. \ref{fig:S2}. Both BESS units were heavily utilized, each completing 0.90 cycles and operating across their full SOC range (15.0\% to 95.0\%), as shown in Fig. \ref{fig:S2_-_b}.

\begin{figure}[h!]
\centering
\includegraphics[width=3.5in]{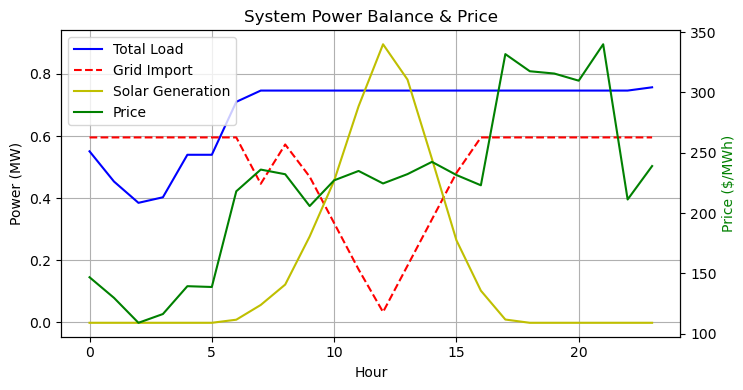}
  \caption{System power and grid price - High Price Day.}
  \label{fig:S3}
\end{figure}
\begin{figure}[h!]
\centering
\includegraphics[width=3.5in]{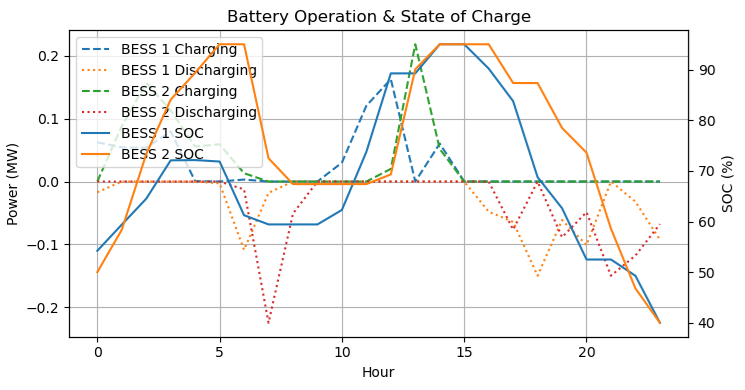}
  \caption{Battery operation and SoC - High Price Day.}
  \label{fig:S3_-_b}
\end{figure}

Economic performance showed substantial improvements, with baseline energy costs of \$812.32 reduced to \$311.90. The peak demand charge of \$1,779.74 led to a total optimized cost of \$2,091.63. Solar generation was abundant at 4.41 MWh, with a near-perfect utilization rate of 100.3\%. Load 10 received the highest solar allocation (0.91 MWh), while Load 9 received the least (0.17 MWh). The Shapley value allocation showed Load 10 receiving the highest solar benefit share (20.6\%) and bearing the highest grid cost share (25.4\%). BESS costs were also highest for Load 10 (18.4\%), reflecting its significant utilization of storage resources. The cooperative operation demonstrated substantial value, with total gains of \$1,711.76 (\$171.18 per participant).

\subsubsection{High Price Day Analysis (March 31, 2015)}

During the high price day, with prices ranging from \$109.06 to \$339.91/MWh, the optimization focused on minimizing expensive grid imports. The system achieved a 21.3\% peak reduction, from 0.757 MW to 0.596 MW, as shown in Fig. \ref{fig:S3}. Both BESS units were moderately utilized, with BESS 1 completing 0.70 cycles and BESS 2 completing 0.84 cycles, as shown in Fig. \ref{fig:S3_-_b}.

The economic impact was significant, with baseline energy costs of \$3,786.07 reduced to \$2,646.09. Despite the peak demand charge of \$5,180.86, the optimization delivered substantial savings. Solar utilization reached 100.4\%, maximizing the value of renewable generation during high-price periods. Load-specific analysis showed Energy Cost variations from \$63.73 (Load 3) to \$449.57 (Load 8), reflecting different consumption patterns and optimization opportunities. The benefit allocation through Shapley values showed Load 8 receiving the highest solar benefit share (18.2\%) but also bearing significant grid costs (16.2\%). The total cooperative gain of \$1,290.92 demonstrated the value of coordinated operation under high price conditions.

\begin{figure}[h!]
\centering
\includegraphics[width=3.5in]{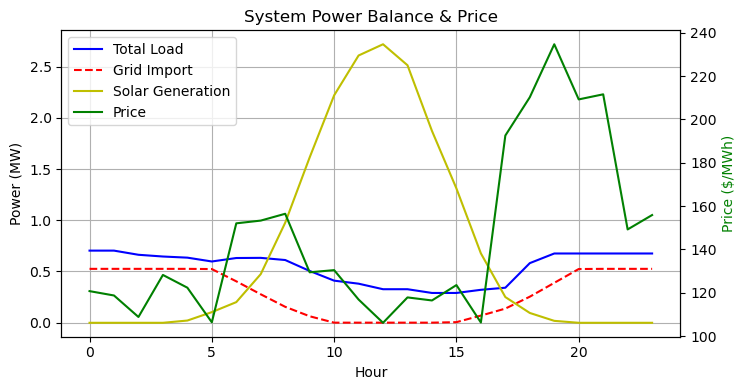}
  \caption{System power and grid price - High Solar Day.}
  \label{fig:S4}
\end{figure}
\begin{figure}[h!]
\centering
\includegraphics[width=3.5in]{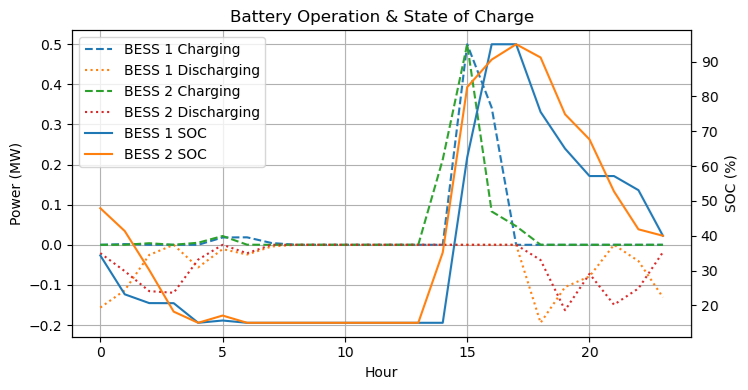}
  \caption{Battery operation and SoC - High Solar Day.}
  \label{fig:S4_-_b}
\end{figure}

\subsubsection{High Solar Day Analysis (July 16, 2015)}

The high solar day scenario presented unique opportunities for maximizing renewable energy utilization. With solar generation reaching 17.66 MWh and peak solar power of 2.72 MW, the system had abundant renewable resources to optimize, as shown in Fig. \ref{fig:S4}. The load profile showed a total energy consumption of 12.97 MWh with a peak power of 0.70 MW, presenting a load factor of 0.77.

The optimization framework achieved a 25.3\% peak reduction, bringing the peak demand down from 0.704 MW to 0.525 MW. This significant reduction was accomplished through coordinated battery operation and strategic solar power allocation. Both BESS units were heavily utilized, as shown in Fig. \ref{fig:S4_-_b}, with BESS 1 completing 0.94 cycles and BESS 2 completing 0.93 cycles. The batteries maintained their SOC within the optimal range of 15.0\% to 95.0\%, demonstrating effective energy time-shifting capabilities. Economic performance showed marked improvement, with baseline energy costs of \$1,930.68 reduced to \$1,066.69. Including the peak demand charge of \$4,571.50, the total optimized cost was \$5,638.19. Notably, the solar utilization rate was 33.1\%, reflecting the challenge of matching abundant solar generation with load requirements. This relatively low utilization rate highlights the potential need for additional storage capacity or load flexibility during high solar generation periods.

\begin{figure}[h!]
\centering
\includegraphics[width=3.5in]{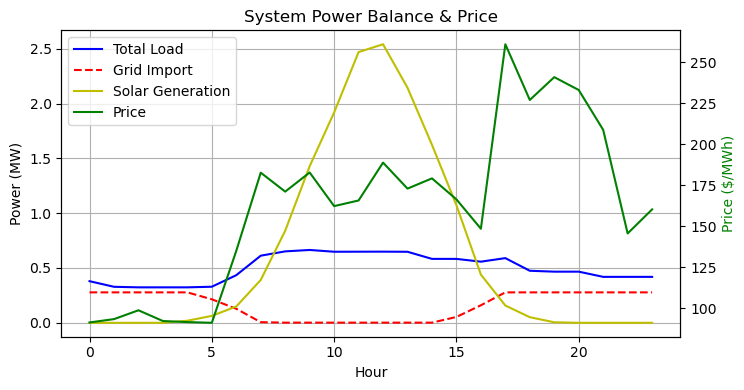}
  \caption{System power and grid price - Typical Weekday.}
  \label{fig:S5}
\end{figure}
\begin{figure}[h!]
\centering
\includegraphics[width=3.5in]{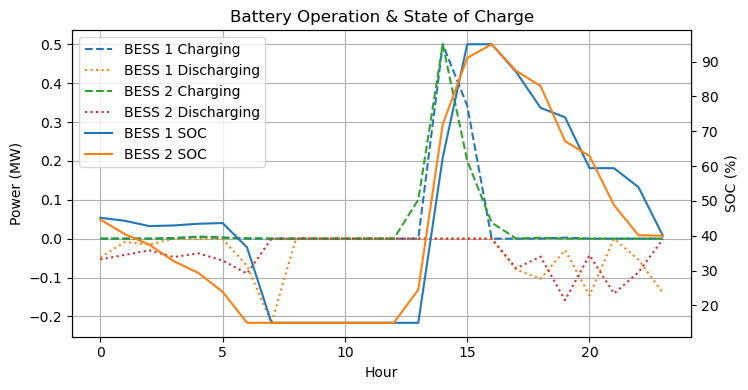}
  \caption{Battery operation and SoC - Typical Weekday.}
  \label{fig:S5_-_b}
\end{figure}

Individual load analysis revealed diverse patterns of resource utilization. Load 10 received the highest solar allocation (1.71 MWh) but showed net battery discharge of -0.51 MWh, indicating strategic energy shifting. In contrast, Load 9 received the lowest solar allocation (0.52 MWh) with a modest battery support of 0.20 MWh. Energy costs varied significantly among loads, from \$15.89 for Load 1 to \$177.73 for Load 7, reflecting different consumption patterns and optimization opportunities. The Shapley value analysis revealed interesting patterns in benefit distribution. Load 10 received the highest solar benefit share (29.3\%) but also bore the highest BESS cost share (36.9\%), reflecting its significant utilization of system resources. The cooperative operation demonstrated substantial value, with total cooperative gains of \$1,801.01, averaging \$180.10 per participant.

\subsubsection{Typical Weekday Analysis (June 3, 2015)}

The typical weekday scenario provided insights into regular operational patterns of the microgrid. With total load energy of 11.92 MWh and solar generation of 15.29 MWh, the system operated with abundant renewable resources. The optimization achieved an impressive 58.3\% peak reduction, reducing the original peak load from 0.663 MW to 0.276 MW, as shown in Fig. \ref{fig:S5}.

\begin{figure}[h!]
\centering
\includegraphics[width=3.5in]{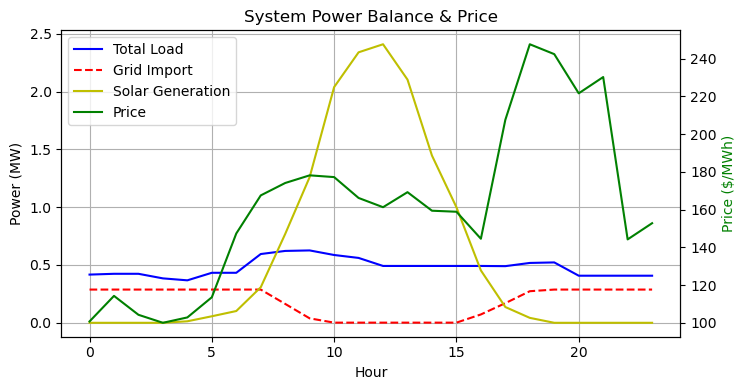}
  \caption{System power and grid price - Typical Weekend.}
  \label{fig:S6}
\end{figure}
\begin{figure}[h!]
\centering
\includegraphics[width=3.5in]{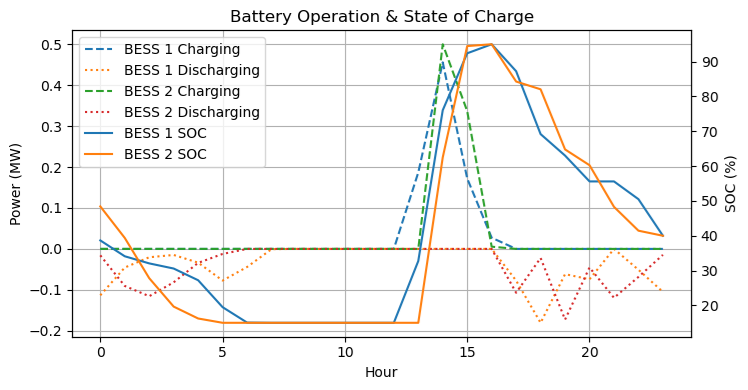}
  \caption{Battery operation and SoC - Typical Weekend.}
  \label{fig:S6_-_b}
\end{figure}

Both BESS units were efficiently utilized, as shown in Fig. \ref{fig:S5_-_b}, each completing 0.91 cycles and maintaining their SOC between 15.0\% and 95.0\%. The economic benefits were substantial, with baseline energy costs of \$2,015.59 reduced to \$606.66. The peak demand charge of \$2,404.83 resulted in a total optimized cost of \$3,011.49. Solar utilization reached 51.8\%, demonstrating effective matching of renewable generation with load requirements during typical operation.  Load-specific analysis showed varied patterns of resource utilization. Load 10 received the highest solar allocation (1.62 MWh) but had net battery discharge of -0.35 MWh. Energy costs ranged from \$15.90 for Load 1 to \$143.54 for Load 7. The Shapley value allocation showed Load 10 receiving the highest solar benefit share (20.5\%) and bearing the highest BESS cost share (20.4\%). The total cooperative gain was \$1,387.52, averaging \$138.75 per participant.

\subsubsection{Typical Weekend Analysis (May 30, 2015)}

The weekend scenario exhibited different consumption patterns compared to weekdays, with total load energy of 11.46 MWh and solar generation of 14.45 MWh, as shown in Fig. \ref{fig:S6}. The system achieved a 54.2\% peak reduction, lowering the peak demand from 0.625 MW to 0.286 MW. Battery utilization remained high, with both BESS units completing 0.90 cycles each, as shown in Fig. \ref{fig:S6_-_b}.

Economic performance showed significant improvement, with baseline energy costs of \$1,887.80 reduced to \$703.34. The peak demand charge of \$2,490.18 led to a total optimized cost of \$3,193.52. Solar utilization reached 47.8\%, slightly lower than weekday utilization due to different load patterns. The benefit allocation showed interesting weekend patterns. Load 5 received the highest solar benefit share (16.4\%) and bore significant BESS costs (18.7\%). The cooperative operation delivered total gains of \$1,704.88, averaging \$170.49 per participant.

\begin{figure*}[t]
\centering
\includegraphics[width=7in]{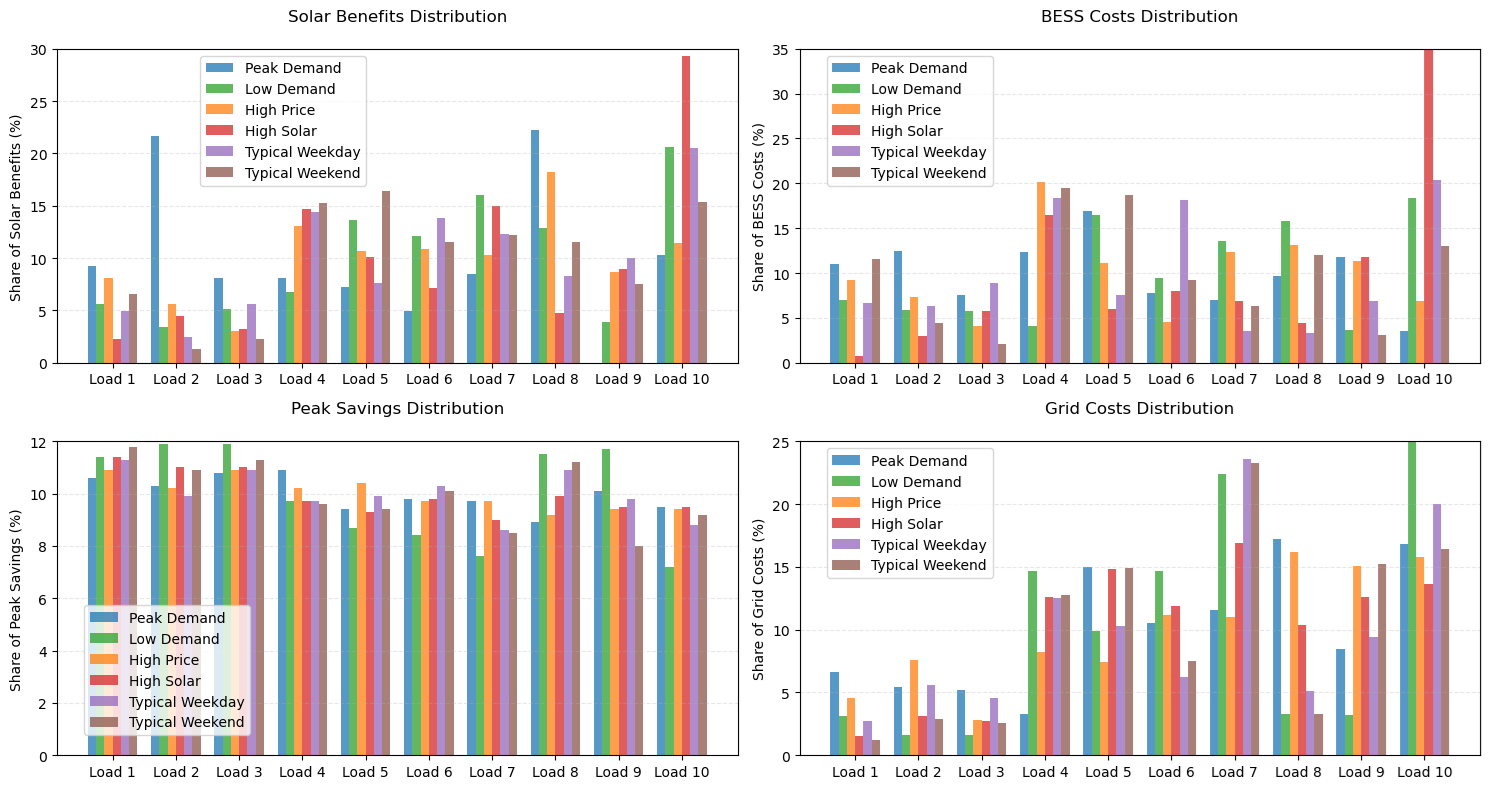}
  \caption{Distribution of costs, savings, and benefits across loads.}
  \label{fig:A}
\end{figure*}

\subsection{Comprehensive Performance Analysis}

To facilitate comprehensive comparison across scenarios, key performance metrics are summarized in Table \ref{tab:scenario_comparison}.

\begin{table}[h]
\caption{Cross-Scenario Performance Comparison}
\label{tab:scenario_comparison}
\setlength{\tabcolsep}{4pt}  
\begin{tabular}{@{}l@{\hspace{8pt}}c@{\hspace{4pt}}c@{\hspace{4pt}}c@{\hspace{4pt}}c@{\hspace{4pt}}c@{\hspace{4pt}}c@{}}
\hline
\multirow{2}{*}{Metric} & Peak & Low & High & High & \multicolumn{2}{c}{Typical} \\
& Demand & Demand & Price & Solar & Weekday & Weekend \\
\hline
\begin{tabular}[t]{@{}l@{}}Peak\\Reduction (\%)\end{tabular} & 7.8 & 62.6 & 21.3 & 25.3 & 58.3 & 54.2 \\[2ex]
\begin{tabular}[t]{@{}l@{}}Solar\\Utilization (\%)\end{tabular} & 114.8 & 100.3 & 100.4 & 33.1 & 51.8 & 47.8 \\[2ex]
\begin{tabular}[t]{@{}l@{}}Average\\BESS Cycles\end{tabular} & 0.34 & 0.90 & 0.77 & 0.94 & 0.91 & 0.90 \\[2ex]
\begin{tabular}[t]{@{}l@{}}Cost\\Reduction (\%)\end{tabular} & 13.0 & 45.0 & 14.2 & 24.2 & 31.5 & 34.8 \\[2ex]
\begin{tabular}[t]{@{}l@{}}Cooperative\\Gain (\$)\end{tabular} & 1,615.72 & 1,711.76 & 1,290.92 & 1,801.01 & 1,387.52 & 1,704.88 \\
\hline
\end{tabular}
\end{table}

\begin{figure*}[t]
\centering
\includegraphics[width=5.5in]{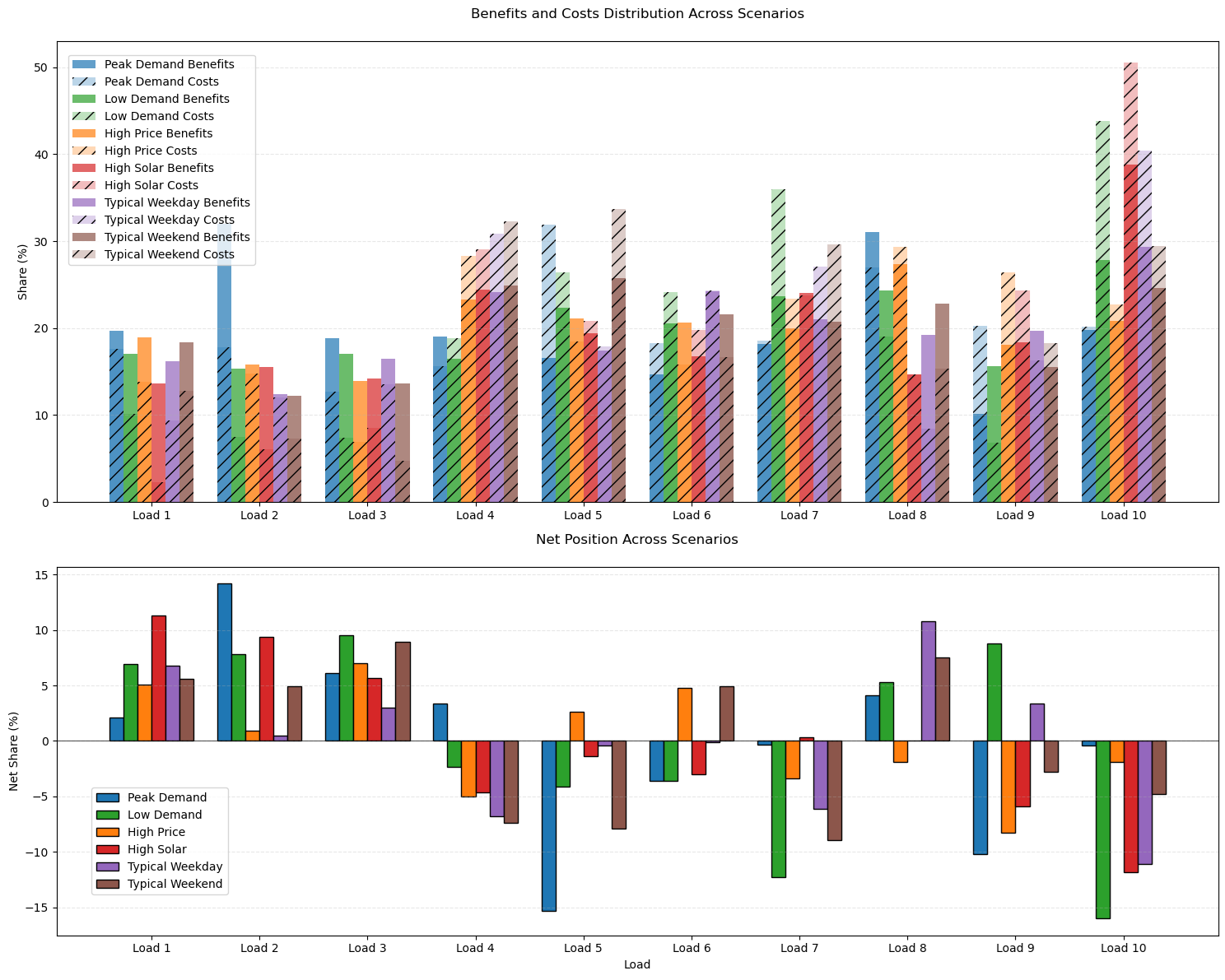}
  \caption{Comparison of costs, benefits, and respective net position across scenarios.}
  \label{fig:B}
\end{figure*}

The economic performance across all scenarios demonstrates consistent cost reductions through optimal resource coordination. Cost savings range from 13.0\% in the peak demand day to 45.0\% in the low demand day scenario. This variation in savings is primarily attributed to different opportunities for load shifting and solar utilization across scenarios. The impact of peak demand charges is particularly significant, contributing between 40\% to 63\% of total costs across scenarios, highlighting the importance of peak reduction strategies. A detailed breakdown of economic performance metrics is presented in Table \ref{tab:economic_metrics}.

\begin{table}[h]
\caption{Economic Performance Metrics Across Scenarios}
\label{tab:economic_metrics}
\begin{tabular}{lcccc}
\hline
Scenario & Baseline & Optimized & Peak & Total \\
 & Energy (\$) & Energy (\$) & Charge (\$) & Cost (\$) \\
\hline
Peak Demand & 4,084.46 & 3,964.31 & 6,840.90 & 10,805.21 \\
Low Demand & 812.32 & 311.90 & 1,779.74 & 2,091.63 \\
High Price & 3,786.07 & 2,646.09 & 5,180.86 & 7,826.96 \\
High Solar & 1,930.68 & 1,066.69 & 4,571.50 & 5,638.19 \\
Typical Weekday & 2,015.59 & 606.66 & 2,404.83 & 3,011.49 \\
Typical Weekend & 1,887.80 & 703.34 & 2,490.18 & 3,193.52 \\
\hline
\end{tabular}
\end{table}

Battery utilization patterns show strategic deployment across different scenarios. The average number of daily cycles ranges from 0.34 during the peak demand day to 0.94 during high solar periods. The consistent maintenance of SOC within operational limits (15-95\%) while achieving significant peak reductions demonstrates the effectiveness of the battery dispatch strategy. Terminal SOC requirements of 40\% were consistently met across all scenarios, ensuring system readiness for subsequent operational cycles.

Solar utilization metrics reveal interesting patterns. The peak demand day shows utilization above 100\% (114.8\%) due to effective energy storage and shifting, while the high solar day shows lower utilization (33.1\%) due to generation exceeding immediate load requirements. This highlights the need for additional storage capacity or load flexibility during high generation periods. 
Peak demand reduction effectiveness varies significantly across scenarios, as shown in Table \ref{tab:technical_metrics}. 

\begin{table}[h]
\caption{Technical Performance Analysis}
\label{tab:technical_metrics}
\setlength{\tabcolsep}{4pt}  
\begin{tabular}{@{}l@{\hspace{6pt}}c@{\hspace{4pt}}c@{\hspace{4pt}}c@{\hspace{4pt}}c@{\hspace{4pt}}c@{}}
\hline
\multirow{2}{*}{Scenario} & Original & Optimized & Peak & BESS & Solar \\
 & Peak (MW) & Peak (MW) & Reduction (\%) & Cycles & Util. (\%) \\
\hline
Peak Demand & 0.853 & 0.786 & 7.8 & 0.34 & 114.8 \\
Low Demand & 0.546 & 0.205 & 62.6 & 0.90 & 100.3 \\
High Price & 0.757 & 0.596 & 21.3 & 0.77 & 100.4 \\
High Solar & 0.704 & 0.525 & 25.3 & 0.94 & 33.1 \\
Typical Weekday & 0.663 & 0.276 & 58.3 & 0.91 & 51.8 \\
Typical Weekend & 0.625 & 0.286 & 54.2 & 0.90 & 47.8 \\
\hline
\end{tabular}
\end{table}

The Shapley value-based cost allocation shown in Fig. \ref{fig:A} demonstrates consistent fairness across scenarios while adapting to varying operational conditions. Analysis of net positions shows that most participants maintain positions within ±10\% of their load proportion, indicating balanced benefit distribution. Notable exceptions occur during high solar periods where some participants achieve higher benefits due to better alignment with generation patterns.

The cooperative gains analysis reveals consistent benefits from coordination across all scenarios, with average per-participant gains ranging from \$129.09 to \$180.10. The stability of these gains across different operational conditions supports the long-term viability of the cooperative arrangement. Table \ref{tab:fairness_metrics} presents the distribution of benefits and costs across loads. Fig. \ref{fig:B} shows the comparison of costs and benefits with respective net position across scenarios.

\begin{table}[h]
\caption{Average Benefit and Cost Distribution Across Loads}
\label{tab:fairness_metrics}
\begin{tabular}{lcccc}
\hline
Load & Solar & BESS & Peak & Grid \\
Category & Benefits (\%) & Costs (\%) & Savings (\%) & Costs (\%) \\
\hline
Small (1-3) & 5-10 & 5-8 & 10-12 & 3-6 \\
Medium (4-6) & 10-15 & 8-12 & 9-11 & 8-12 \\
Large (7-10) & 15-25 & 12-20 & 8-10 & 12-20 \\
\hline
\end{tabular}
\end{table}

The comprehensive analysis reveals significant achievements in both economic and technical performance metrics. The framework delivers consistent cost reductions ranging from 13.0\% to 45.0\% across all scenarios, with peak demand reductions reaching up to 62.6\% in low demand conditions. Solar utilization rates exceed 100\% during peak periods through effective energy shifting, while BESS units demonstrate adaptive cycling patterns that maintain efficient operation within lifecycle constraints. The Shapley value-based allocation mechanism successfully balances individual contributions and benefits, ensuring participant satisfaction across varying operational conditions, with stable cooperative gains supporting long-term arrangement viability. Moreover, the framework's consistent performance across diverse operational scenarios and maintained computational efficiency through strategic problem formulation suggests excellent scalability potential for larger systems, validating the approach's effectiveness in achieving both technical optimization and fair benefit distribution in practical microgrid operations.

\section{Conclusion}
\label{ch:conclusion}

The proposed framework successfully addresses the dual challenges of optimal microgrid operation and equitable cost allocation through a comprehensive integration of optimization and game theory approaches. The mixed-integer linear programming model effectively manages the complex interactions between various system components, achieving substantial peak reductions and maximizing renewable utilization across diverse operational scenarios. The Shapley value-based allocation mechanism ensures fair distribution of benefits and costs, promoting long-term participant satisfaction and system sustainability. The framework's effectiveness is validated through extensive testing across six distinct scenarios, including peak demand, low demand, high price, high solar generation, and typical operating conditions. The results demonstrate consistent performance improvements, with peak demand reductions of up to 62.6\% during low demand periods and effective solar utilization reaching 114.8\% through optimal storage coordination. The fairness of the allocation mechanism is confirmed by balanced net positions across participants, with large loads achieving benefit shares proportional to their contributions and smaller loads receiving equitable compensation for their flexibility. Future work could explore dynamic adaptation of the framework to changing participant preferences, integration of additional distributed energy resources, and extension to larger-scale community energy systems. The framework's modular design allows for straightforward incorporation of new optimization objectives and fairness criteria as community needs evolve.

\bibliographystyle{IEEEtran}
\bibliography{main}

\end{document}